\documentclass[
 reprint,
 amsmath,amssymb,
 aps,
 pra,
]{revtex4-2}

\usepackage{graphicx}
\usepackage{dcolumn}
\usepackage{bm}
\usepackage{hyperref}

\begin{document}

\preprint{APS/123-QED}

\title{Sending or not sending twin-field quantum key distribution\\
 with distinguishable decoy states}

\author{Yi-Fei Lu}
\author{Mu-Sheng Jiang}
 \email{jms@qiclab.cn}
\author{Yang Wang}
\author{Xiao-Xu Zhang}
\author{Fan Liu}
\author{Chun Zhou}
\author{Hong-Wei Li}
\author{Wan-Su Bao}
 \email{bws@qiclab.cn}
\affiliation{Henan Key Laboratory of Quantum Information and Cryptography, SSF IEU, Zhengzhou, Henan 450001, China\\
Synergetic Innovation Center of Quantum Information and Quantum Physics, University of Science and Technology of China, Hefei, Anhui 230026, China
}

\date{\today}

\begin{abstract}
Twin-field quantum key distribution (TF-QKD) and its variants can overcome the fundamental rate-distance limit of QKD which has been demonstrated in the laboratory and field while their physical implementations with side channels remains to be further researched. We find the external modulation of different intensity states through the test, required in those TF-QKD with post-phase compensation, shows a side channel in frequency domain. Based on this, we propose a complete and undetected eavesdropping attack, named passive frequency shift attack, on sending or not-sending (SNS) TF-QKD protocol given any difference between signal and decoy states in frequency domain which can be extended to other imperfections with distinguishable decoy states. We analyze this attack by giving the formula of upper bound of real secure key rate and comparing it with lower bound of secret key rate under Alice and Bob's estimation with the consideration of actively odd-parity pairing (AOPP) method and finite key effects. The simulation results show that Eve can get full information about the secret key bits without being detected at long distance. Our results emphasize the importance of practical security at source and might provide a valuable reference for the practical implementation of TF-QKD.

\end{abstract}

\maketitle

\section{\label{intro}Introduction}

Quantum key distribution (QKD) allows two distant parties, Alice and Bob, to share secret keys securely in the presence of an eavesdropper, Eve, by harnessing the laws of physics \cite{bennett2014RN153,shor2000RN300,xu2020RN130}. Combined with one-time pad, Alice and Bob can achieve unconditionally secure private communication. Notable progress has been made to improve performance, such as the communication distance and secret key rate, and bridge the gap between the idealized device models assumed in security proofs and the functioning of realistic devices in practical systems.

The measurement-device-independent (MDI) QKD \cite{lo2012RN72} can remove both known and unknown security loopholes, or so-called side channels, in the measurement unit perfectly which shift the focus of quantum attacks to the source. Photon-number-splitting (PNS) attack \cite{brassard2000RN299,lutkenhaus2002RN301}, the major threat at source since single-photon source is not available at present and weak laser light are widely used in practical QKD systems, has been overcome by the decoy-state method \cite{hwang2003RN202,lo2005RN81}. Combining these two methods, the decoy-state MDI-QKD equipped with some security patches performs well with imperfect single-photon sources. However, the key rate and communication distance are two implementation bottlenecks. To exceed the linear scale of key rate \cite{takeoka2014RN231,pirandola2017RN103}, twin-field QKD (TF-QKD) was proposed by Lucamarini \emph{et al.} \cite{lucamarini2018RN45} whose key rate scales linearly with square-root of the channel transmittance $\eta$ by harnessing single-photon interference over long distance. Though the security is not completed and the security loophole is caused by the later announcement of the phase information \cite{wang2018RN22}, then many variants of TF-QKD \cite{wang2018RN22,ma2018RN56,curty2019RN57,cui2019RN41,lin2018RN391,tamaki2018RN280} have been proposed to deal with this security loophole and each has its advantages. Many effects have been considered in real-life implementation to accelerate its application, including finite key effects, the number of states with different intensities, the phase slice of appropriate and asymmetric transmission distance, etc. \cite{yu2019RN24,jiang2019RN16,xu2020RN23,jiang2020RN17,hu2019RN15,zhou2019RN26,yin2019RN333,zeng2020RN388,curras2021RN416,zhang2020RN382,grasselli2019RN42,teng2020RN48,lu2019RN408,wang2020RN409,wang2020RN410,lorenzo2019RN412,mao2021RN414}. Meanwhile, several experiments of TF-QKD have been completed in laboratory and field to demonstrate its ability to overcome the rate-distance limit \cite{minder2019RN46,wang2019RN284,zhong2019RN52,liu2019RN19,chen2020RN13,fang2020RN55,liu2021RN395,chen2021RN418}.

However, the physical implementations of TF-QKD protocols with side channels remains to be further researched at present. Since TF-QKD retains the MDI characteristic, we should just focus on light source. Ideally, it is assumed that the sending devices are placed in a protected laboratory, and can prepare and encode quantum states correctly. Unfortunately, these conditions are not met in practical systems, and state preparation flaws (SPFs) and leakage may be induced from imperfect devices or Eve's disturbance. A small imperfection at source does not necessarily mean a small impact on the secret key rate, because Eve could enhance such imperfection by exploiting channel loss. Therefore, Eve can steal secret information actively by performing Trojan-horse attack \cite{vakhitov2001RN194,gisin2006RN187,fung2007RN198,jiang2012RN180,jiang2014RN181,jain2014RN189,lucamarini2015RN191,bugge2014RN318,sun2015RN320,makarov2016RN343,huang2019RN188,huang2020RN312,pang2020RN192} on modulators, lasers or attenuators and so on, or passively by harnessing the SPFs and leakage caused by imperfect devices \cite{brassard2000RN299,lutkenhaus2002RN301,tang2013RN305,tamaki2016RN304,huang2018RN303,sajeed2015RN313}.

In those QKD protocols with imperfect single-photon sources, the decoy-state method is vital and employed to monitor the channel eavesdropping whose security is based on the fact that Eve could not distinguish between decoy and signal states. In practice, however, it does not with the behavior of the real apparatuses or Eve's disturbance. For instance, the probability distributions of signal states and decoy states do not overlap in time domain totally with pump-current modulation \cite{huang2018RN303}. Besides, signal states and decoy states can be distinguished with external intensity modulation in frequency domain when Eve applying wavelength-selected photon-number-splitting attack actively in 'plug-and-play' systems \cite{jiang2012RN180}. This loophole is introduced by decoy-state method and caused by the imperfections of modulators and modulation voltage. As the decoy-state method is employed in TF-QKD protocols, it is of significance to analyze its practical security in this aspect.

In this paper, we concentrate on the sending or not-sending (SNS) TF-QKD protocol \cite{wang2018RN22} with actively odd-parity pairing (AOPP) method and finite-key effects taken into consideration \cite{xu2020RN23,jiang2020RN17}, and propose a complete and undetected eavesdropping attack, named passive frequency shift attack, which can take advantage of the most general side channels in frequency domain and can be extended to other cases with distinguishable decoy states. In Sec. \ref{passi}, we recap the frequency shift of intensity modulators (IMs) and test experimentally the spectral distribution of signal pulses with external modulation method which shows a side channel in frequency domain. In Sec. \ref{attack}, we propose a passive frequency shift attack on SNS protocol and analyze its adverse impact by giving the formula of upper bound of secure key rate and comparing it with the lower bound of secret key rate under Alice and Bob's estimation with consideration of AOPP method and finite-key effects. In Sec. \ref{numer}, we present our simulation results. Last, we give some discussion about the countermeasure of side channels in Sec. \ref{diss} and conclude in Sec. \ref{conl}.

\section{\label{passi}Frequency shift of intensity modulators}

In this section, we will recap the frequency shift of IMs and test experimentally to show a side channel in frequency domain.

There are several kinds of intensity modulators, such as the Mach-Zehnder type electro-optical intensity modulators (EOIMs), electro-absorption modulators (EAMs), and acousto-optical modulators (AOMs). EOIMs, especially LiNbO$_3$-based devices, possess the excellent performance of wavelength-independent modulation characteristics, excellent extinction performance (typically 20 dB) and low insertion losses (typically 5 dB) \cite{winzer2006RN279}.

LiNbO$_3$-based EOIMs work by the principle of interference, controlled by modulating the optical phase. The incoming light is coupled into a waveguide and then split into two paths of a Mach-Zehnder interferometer equally and interfere at an output coupler. The two arms made of lithium niobate will induce a phase change when applying voltages. Accordingly, the intensity and phase of output light will be modulated after interference depending on the applied electrical voltages. Assuming voltages $V_1(t)$ and $V_2(t)$ are applied to two arms separately with the input field of intensity $E_0$ and frequency $\omega_0$, the output field can be written as \cite{jiang2012RN180}
\begin{eqnarray}
    E_{\rm out}(t) = E_0 {\rm cos} [\Delta \varphi(t)] e^{i[\omega_0 t+\varphi(t)]},
\end{eqnarray}
where $\Delta \varphi(t)= [\gamma V_1(t) +\varphi_{1} - \gamma V_2(t)-\varphi_{2}]/2$ and $\varphi(t)= [\gamma V_1(t) +\varphi_{1} + \gamma V_2(t)+\varphi_{2}]/2$, $\gamma = \pi/V_{\pi}$ is the voltage-to-phase conversion cofficients for two arms, and $\varphi_{1}$ and $\varphi_{2}$ are the static phases which we will omit for simplicity. Here, $V_\pi$ is half-wave voltage that is required to change the phase in one modulator arm by $\pi$ radians. The output intensity is given by
\begin{eqnarray}
    P_{\rm out}(t) = |E_{\rm out}(t)|^2= \frac{P_0}{2} \bigl[ 1+{\rm cos} [ \gamma V(t) ] \bigr],
    \label{eq3}
\end{eqnarray}
where $V(t)=V_1(t)-V_2(t)$ and $P_0$ is the input optical power. The phase maintains and intensity is determined by Eq.~\ref{eq3} on condition that the two modulator arms are driven by the same amount, but in opposite directions (i.e. $V_1 (t)=-V_2 (t)$), which is known as balanced driving or push–pull operation. When $V(t)$ is constant we will get pure intensity modulation without frequency shift. However, once $V(t)$ is not a constant any more, something unexpected will arise to the output field. Specifically, frequency shift will be induced as we can see, for example, if $V_1(t)=-V_2(t)=V_0+kt$, the output field can be expressed as \cite{jiang2012RN180}
\begin{eqnarray}
    E_{\rm out} = \frac{E_0}{2} \Bigl[ e^{i[(\omega_0 + \gamma k)t + \gamma V_0 ] }  + e^{i[(\omega_0 - \gamma k)t + \gamma V_0 ] } \Bigr].
    \label{eq4}
\end{eqnarray}
From Eq.~\ref{eq4}, we can see a frequency shift of the light pulses with $\pm \omega_m =\pm \gamma k$ compared with the original $\omega_0$, where $k$ is the slope of modulation voltage. Moreover, the frequency shift of the output field will be more confusing when the modulation voltages are more complicated in practical systems. We can analyze the spectrum of the output field using the fast Fourier transform method.

\begin{figure}[b]
  \includegraphics[width=0.43\textwidth]{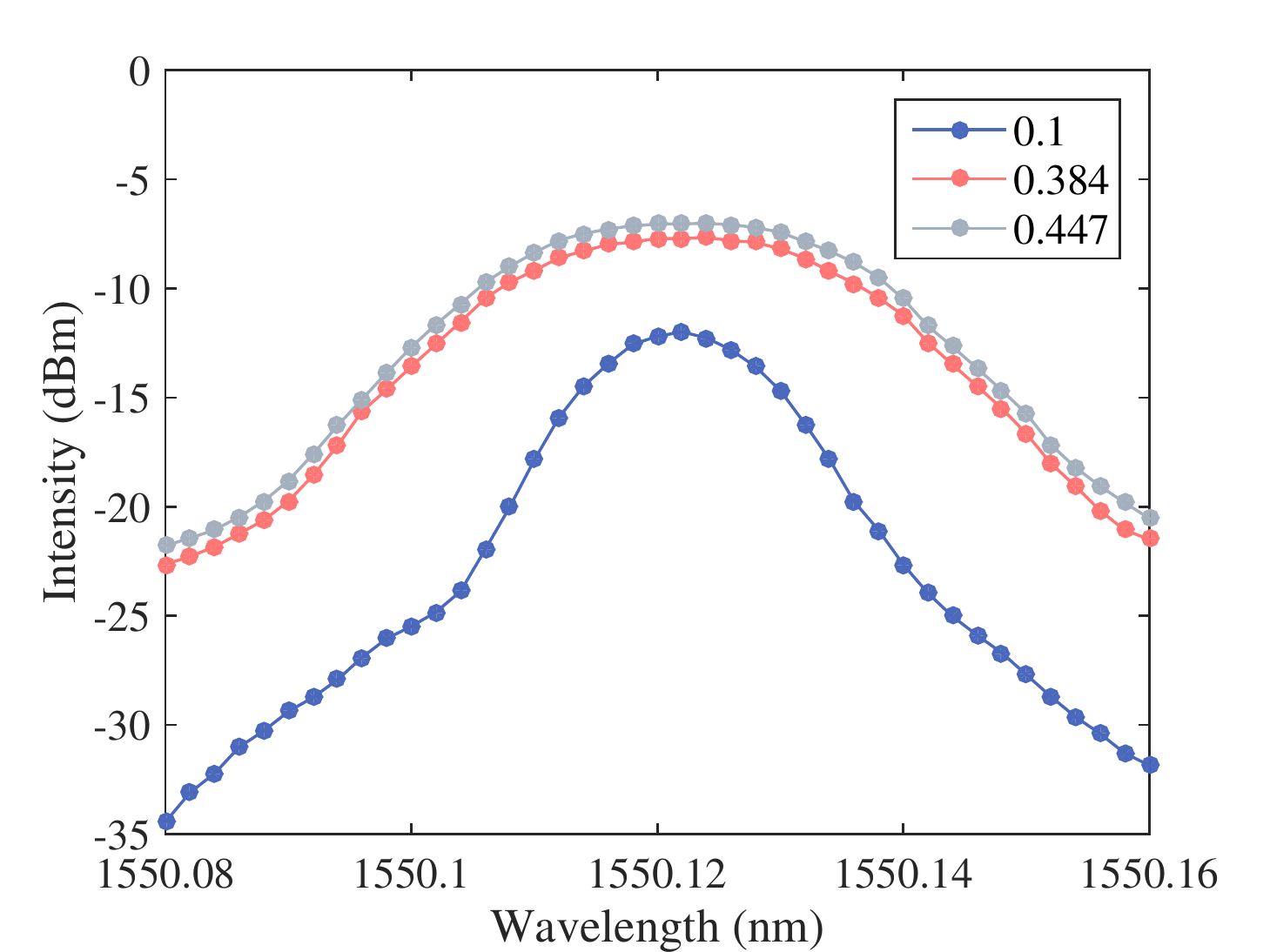}
  \caption{\label{fig1} The wavelength spectrum of three different states with intensity ratio as 0.1: 0.384: 0.447 ($\mu_a=0.1$, $\mu_b=0.384$, $\mu_z=0.447$).}
\end{figure}

\begin{figure}[b]
  \includegraphics[width=0.43\textwidth]{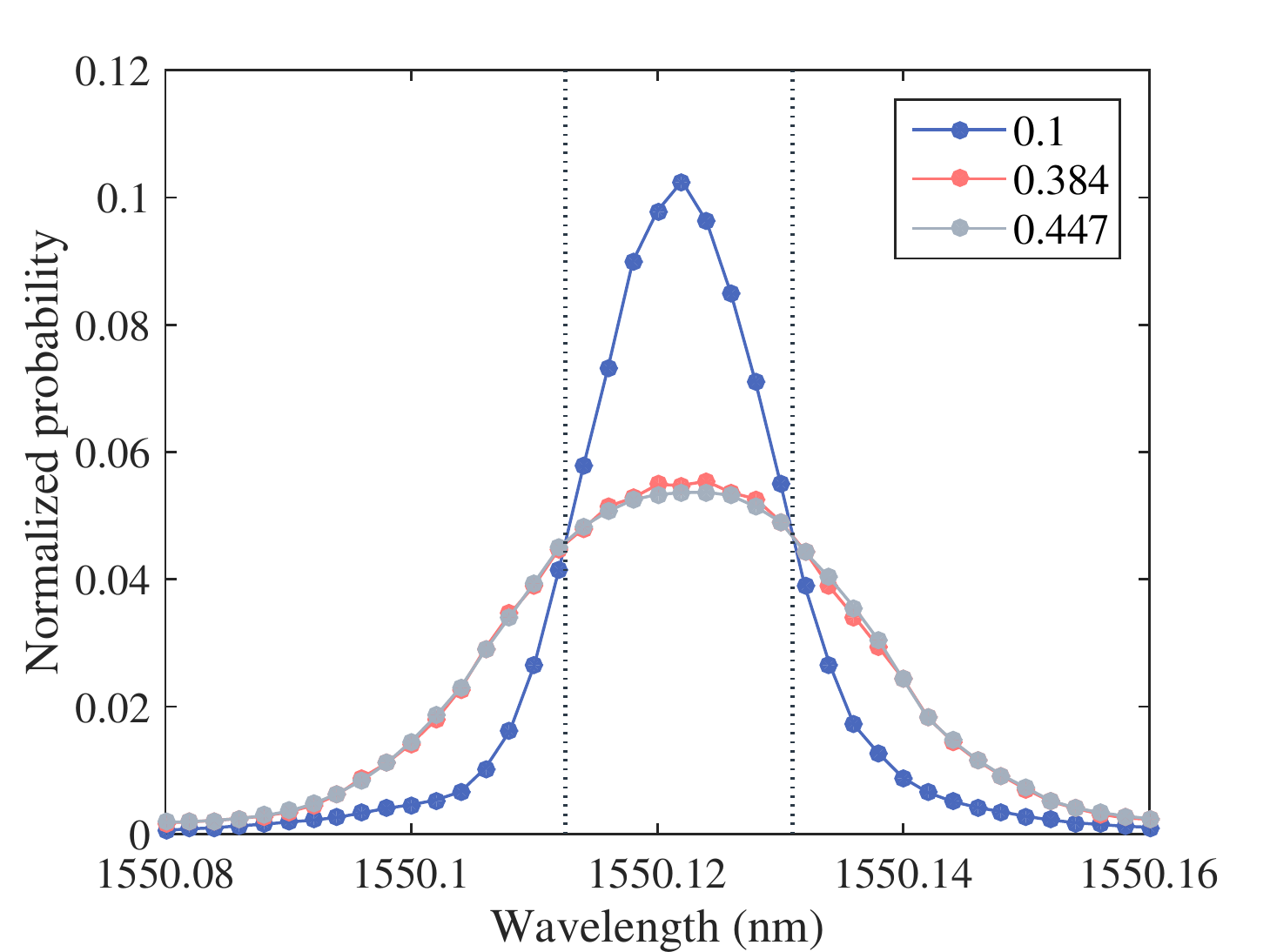}
  \caption{\label{fig2} The normalized intensity distribution of three different states with intensity ratio as 0.1: 0.384: 0.447 ($\mu_a=0.1$, $\mu_b=0.384$, $\mu_z=0.447$). Two dashed lines at 1550.1125 nm and 1550.131 nm are boundaries that can be used to distinguish states with intensity $\mu_z$ ($\mu_b$) and $\mu_a$. There are also some subtle differences between states $\mu_z$ and $\mu_b$ in frequency domain.}
\end{figure}

To evaluate the frequency shift of different intensity pulses, we test it in principle. Optical pulses with 1 ns pulse width are produced with a constant intensity from a laser diode (Keysight 8164B) first, modulated by an IM driven by an arbitrary waveform generator (Keysight M9505A) and measured by an optical spectrum analyzer (Yokogawa AQ637D) last. Note that the measurement is taken before the fixed attenuation as implemented in actual systems since the probability of emitting pulses at the single-photon level follows the same distribution. Fig.~\ref{fig1} illustrates the wavelength spectrum of three states with intensity ratio taken from the SNS experiment \cite{chen2020RN13} as 0.1: 0.384: 0.447 ($\mu_a=0.1$, $\mu_b=0.384$, $\mu_z=0.447$). And the normalized intensity probability distributions are shown in Fig.~\ref{fig2} to distinguish the difference.

Obviously, the states modulated by IMs of different intensities do not overlap totally in frequency domain. The distinction of signal states (also strong decoy states) and weak decoy states is evident as expected because the amplitude of modulation voltages of signal and strong decoy states are higher which will induce more frequency shift, and thus the peaks of the signal and strong decoy states are lower than weak decoy states. More precisely, the normalized probability of weak decoy states is higher than signal and strong decoy states between 1550.112 nm and 1550.131 nm. There are also slight differences between signal and strong decoy states. The difference will be bigger when narrowing pulses or the line width as TF-QKD requires, or when the intensity difference of the different states becomes larger. On this foundation, Eve can apply a passive frequency shift attack by harnessing this side channel.

\section{\label{attack}Passive frequency shift attack on SNS protocol}

In this section, we propose a passive frequency shift attack scheme on practical SNS TF-QKD systems by exploiting the side channels in frequency domain and analyze its adverse impact by fiving the formula of upper bound of secure key rate with consideration of AOPP method and finite-key effects. But we note that this attack can be applied with other side channels except in frequency domain.

In fact, the signal pulses (including signal states and decoy states) and reference pulses must be modulated with a stable continuous-wave laser source with external modulation method so as to estimate and compensate phase noise in those TF-QKD protocols which need post-phase compensation, such as SNS TF-QKD \cite{wang2018RN22} and phase-matching (PM) TF-QKD \cite{ma2018RN56}. Even if the synchronization can not be controlled by Eve unlike discussed in Ref. \cite{jiang2012RN180,jiang2014RN181}, the probability distributions of signal states and decoy states may do not overlap inevitably in frequency domain.

In the 4 intensity SNS TF-QKD protocol (see Appendix \ref{proto} for details), Alice and Bob need to modulate the continuous light to 5 different intensities, the maximum intensity pulses are used as phase reference pulses, the minimum as vacuum states, while others as signal states, weak and strong decoy states. In practical SNS systems \cite{liu2019RN19,chen2020RN13}, three IMs are used to modulate these 5 different pulses to ensure that the output signal intensities are in agreement with the theoretical requirements and the reference detections are high enough for phase compensation. There will be side channels in frequency domain inevitably. In what follows, we introduce the passive frequency shift attack. 

\begin{figure}[ht]
  \includegraphics[width=0.45\textwidth]{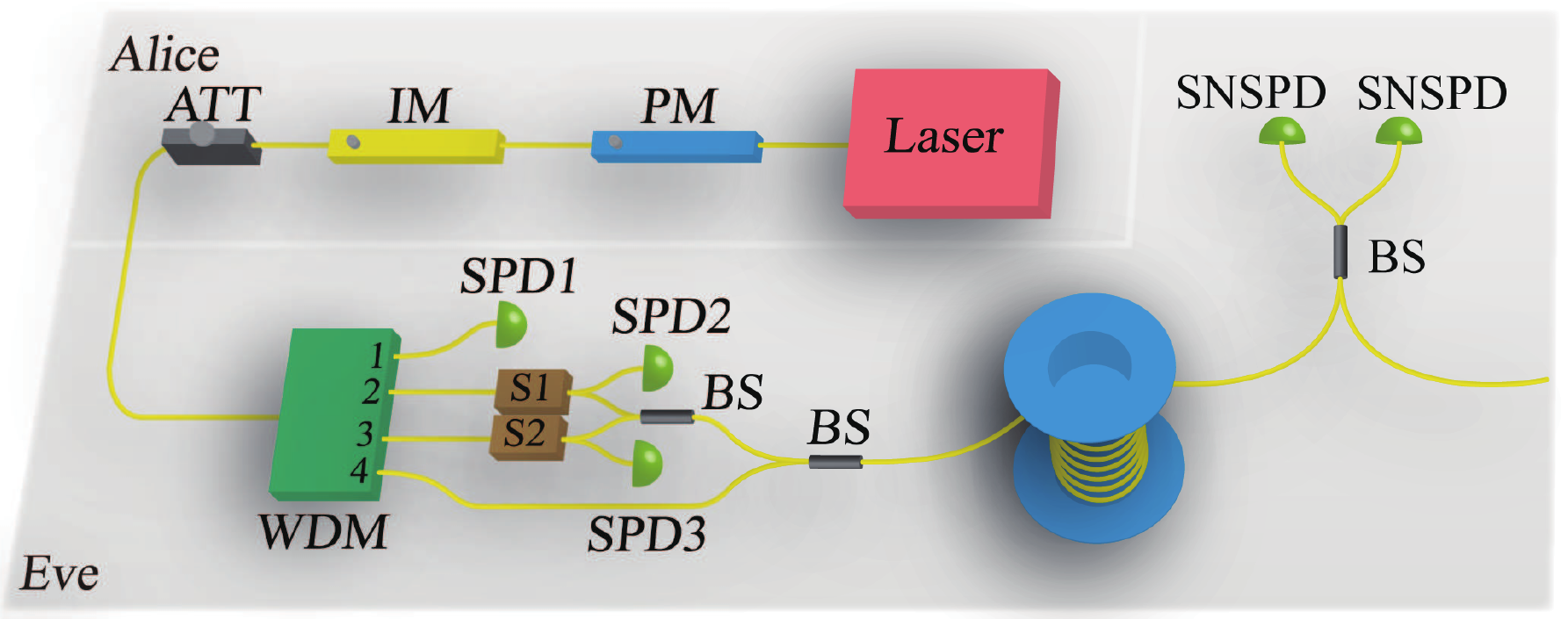}
  \caption{\label{fig3} Schematic of the passive frequency shift attack. PM: phase modulator, IM: intensity modulator, ATT: attenuator, WDM: wavelength division multiplexer, SPD: single-photon detector, S: light path selector, BS: beam splitter, SNSPD:  superconducting nanowire single-photon detectors. The light path selector S1 (S2) is controlled by SPD1 (SPD1 and SPD2) and Bob's device is the same as Alice's  which we have omitted in figure.}
\end{figure}

Suppose Eve intercepts all signal pulses at Alice and Bob's output ports where the signal pulses haven't been attenuated by channels, and then distinguishes signal and decoy states with a wavelength division multiplexer (WDM) and three single-photon detectors (SPDs) as illustrated in Fig.~\ref{fig3}. Set internals $T_\alpha$ properly, where $\alpha \in \{z,a,b\}$, according to the wavelength spectrum of different states to distinguish the states with intensity $\mu_\alpha$ but may fail with a certain probability. Theoretically, $T_\alpha$ can be set as the union of two symmetric intervals according to Eq.~\ref{eq4}.

Suppose the four ports of WDM 1, 2, 3 and 4 can export photons with frequency located in $T_z$, $T_a$, $T_b$ and others. The light path selector S1 (S2) is controlled by SPD1 (SPD1 and SPD2). Denote as 1 or 0 when SPD (i.e. SPD1, SPD2 or SPD3) clicks or not, and 1 or 0 when the light path selector (i.e. S1 or S2) selects the up or down path. Then we set
\begin{eqnarray}
    {\rm S1} &&= \overline{ {\rm SPD1} },\nonumber\\
    {\rm S2} &&= {\rm SPD1} \vee {\rm SPD2}.
\end{eqnarray}
Note that only one SPD at most will click under this principle. According to the response of SPDs, set the total transmittance as $\eta_z$, $\eta_a$, $\eta_b$ or $\eta_k$ when SPD1, SPD2, SPD3 clicks or none of them click, respectively. In this process, Eve could get partial raw key bits after Alice (Bob) announces their signal and decoy windows, which can be understood in this way that Eve can conclude their key bits as 1 (0) when $ {\rm SPD1} \vee {\rm SPD2} \vee {\rm SPD3} = 1 $ in $Z$ windows. There is no bit-flip error between Alice (Bob) and Eve because Eve can intercepts photons at output ports without stray photons. Only the raw bits are secure in $Z$ windows when $ {\rm SPD1} \vee {\rm SPD2} \vee {\rm SPD3} = 0 $ on both sides. On the one hand, once Eve detects photons successfully on one side, the bit is either the same or a bit-flip error with the other side which will be revealed in the error correction step (and pre-error correction process when AOPP is performed). On the other hand, the bits are balanced (i.e. random for Eve) in one-detector heralded events with $ {\rm SPD1} \vee {\rm SPD2} \vee {\rm SPD3} = 0 $ which means the raw bits are secure in these windows. Though Eve could not distinguish the decoy states and signal states without errors, once the transmittance of signal and decoy states differ from others, the decoy-state method may not estimate the count rate and phase-flip error rate of single-photon states correctly. When the actual secure key rate is lower than the estimated one, the final secret string is insecure partially. 

We emphasize that this attack will not introduce unnecessary errors as the beam splitting and measurement by Eve can be seen as a loss and the rest of the pulses are coherent states with no phase noise. What is more, Eve could control errors completely expect the inherent errors of the protocol through channels, superconducting nanowire single-photon detectors (SNSPDs) and classic information he announces. In the following, we will analyze the negative effect of this passive frequency shift attack. 

Consider the most general case, we assume that the envelope of wavelength spectrum can be written as $f_\beta(\lambda)$, where $\beta \in \{ z,a,b \}$. The wavelength spectrum of signal states and decoy states could not totally overlap. By setting the internals $T_\alpha$ properly, Eve can distinguish the decoy states and siganl states with errors. The proportion of state $\mu_\beta$ in internals $T_\alpha$ can be shown as
\begin{eqnarray}
  r_{\beta|\alpha} = \int_{T_\alpha} f_\beta(\lambda) d\lambda,
\end{eqnarray}
where $\alpha$, $\beta \in \{ z,a,b \}$. Which one detector will click obeys a certain probability distribution. Therefore, the states of intensity $\mu_\beta$ would be transformed with one of four different transmittance $\varOmega_\beta = \{\eta_{\beta z}, \eta_{\beta a}, \eta_{\beta b}, \eta_{\beta k}\}$ with a finite probability, where
\begin{eqnarray}
    \eta_{\beta z}=&&\eta_{z} (1-r_{\beta|z}),\nonumber\\
    \eta_{\beta a}=&&\eta_{a} (1-r_{\beta|z}-r_{\beta|a}),\nonumber\\
    \eta_{\beta b}=&&\eta_{b} (1-r_{\beta|z}-r_{\beta|a}-r_{\beta|b}),\nonumber\\
    \eta_{\beta k}=&&\eta_{k} (1-r_{\beta|z}-r_{\beta|a}-r_{\beta|b}).
\end{eqnarray}

\begin{table}[b]
  \caption{\label{tab1} List of experimental parameters. Here, $\gamma$ is the fiber loss coefficient (dB/km), $\eta_d$ is the detection efficiency of detectors, $e_d$ is the misalignment-error probability, $f_{\rm EC}$ is the error correction inefficiency, $\xi$ is the failure probability of statiscal fluctuations analysis, $p_d$ is the dark count rate and $M$ is the number of phase slices.}
  \begin{ruledtabular}
  \begin{tabular}{ccccccc}
  \textrm{$\gamma$}&
  \textrm{$\eta_d$}&
  \textrm{$e_d$}&
  \textrm{$f_{\rm EC}$}&
  \textrm{$\xi$}&
  \textrm{$p_d$}&
  \textrm{$M$}\\
  \colrule
  0.2 & 56\% & 0.1 & 1.1 & $2.2\times 10^{-9}$ & $10^{-10}$ & 16\\
  \end{tabular}
  \end{ruledtabular}
  \end{table}
  
  \begin{table}[b]
  \caption{\label{tab2} List of experimental parameters about intensity and probability Alice and Bob select.}
  \begin{ruledtabular}
  \begin{tabular}{ccccccc}
  \textrm{$\mu_a$}&
  \textrm{$\mu_b$}&
  \textrm{$\mu_z$}&
  \textrm{$p_z$}&
  \textrm{$p_a$}&
  \textrm{$p_b$}&
  \textrm{$p_{z0}$}\\
  \colrule
  0.1 & 0.384 & 0.447 & 0.776 & 0.85 & 0.073 & 0.732\\
  \end{tabular}
  \end{ruledtabular}
  \end{table}
  
  \begin{table}[b]
  \caption{\label{tab3} Five groups of the proportion $r_{\beta|\alpha}$ of state $\mu_\beta$ in internals $T_\alpha$.}
  \begin{ruledtabular}
  \begin{tabular}{cccccccccc}
  \textrm{~}&
  \textrm{$r_{z|z}$}&
  \textrm{$r_{a|z}$}&
  \textrm{$r_{b|z}$}&
  \textrm{$r_{z|a}$}&
  \textrm{$r_{a|a}$}&
  \textrm{$r_{b|a}$}&
  \textrm{$r_{z|b}$}&
  \textrm{$r_{a|b}$}&
  \textrm{$r_{b|b}$}\\
  \colrule
  G1 &0.01 & 0.008 & 0.01 & 0.01  & 0.1 & 0.01  & 0.01 & 0.008 & 0.01 \\
  G2 &0.01 & 0.008 & 0.01 & 0.005 & 0.1 & 0.005 & 0.01 & 0.008 & 0.01 \\
  G3 &0.01 & 0.008 & 0.01 & 0.01  & 0.2 & 0.01  & 0.01 & 0.008 & 0.01 \\
  G4 &0.01 & 0.005 & 0.01 & 0.005 & 0.2 & 0.005 & 0.01 & 0.005 & 0.01 \\
  G5 &0.01 & 0.008 & 0.01 & 0     & 0.1 & 0     & 0.01 & 0.008 & 0.01 \\
  \end{tabular}
  \end{ruledtabular}
  \end{table}

Here, the attenuation coefficient comes from Eve's interfection and detection, and the total transmittance can be controlled by Eve completely which means that Eve is allowed to use a lower-loss or even lossless channel and perfect detectors with 100\% detection efficiency and no dark count, and could select internals freely to obtain a satisfactory results. For states with intensity $\mu_\beta$, the probability of being transmitted with $\eta_{\beta\gamma} \in \varOmega_\beta$, $\gamma \in M = \{z,a,b,k\}$ can be shown as
\begin{eqnarray}
    p_{\beta z} &&=  (1-e^{-\mu_{\beta|z}}), \nonumber\\
    p_{\beta a} &&= e^{-\mu_{\beta|z}} (1-e^{-\mu_{\beta|a}}), \nonumber\\
    p_{\beta b} &&= e^{-\mu_{\beta|z}} e^{-\mu_{\beta|a}}  (1-e^{-\mu_{\beta|b}}), \nonumber\\
    p_{\beta k} &&= e^{-\mu_{\beta|z}} e^{-\mu_{\beta|a}} e^{-\mu_{\beta|b}},
\end{eqnarray}
where $\mu_{\beta|\alpha} = \mu_\beta r_{\beta|\alpha}$. Here, $e^{-\mu_{\beta|\alpha}}$ is the probatility of zero photon in internals $T_\alpha$ with intensity $\mu_\beta$.

Since TF-QKD protocols are proposed for the implementation of long optical fiber communications, Eve's best target is to acquire more percentage of keys by minimizing $\eta_k$ as far as possible while maintaining the key rate and communication distance under Alice and Bob's estimation. When the communication distance is long enough, Eve may steal all secret key bits.

There are two key rates matter: the lower bound of secret key rate under Alice and Bob's estimation $R_e$ and the upper bound of real secure key rate $R_u$. Note that Alice and Bob could not estimate $R_e$ correctly under this attack since it is impossible to pick out the decoy states that have undergone the same operation as the signal states, i.e. the decoy-state method does not work properly.

When AOPP method is performed with partial bits leaked to Eve, Bob only chooses odd-parity bit pairs and they will keep the second bit if Alice's bits pairs are odd, too. In this process, Eve's information on raw bits does not reduce as the parity information is public. Therefore, the upper bound of secure key rate $R_u$ can be shown as
\begin{eqnarray}
  R_u = n_{1,{\rm sec}} [ 1 -H( e_{\rm in} ) ] - n_t' f H(E_Z'),
\end{eqnarray}
where $n_{1,{\rm sec}}$ is the upper bound of single photons which can be used to distill secure key bits and $e_{\rm in}$ is the inherent phase-flip error rate (i.e. the lower bound of untagged bits) determined by the number of phase slices $M$. These two parameters will be discussed in Appendix \ref{detail}.

This attack can be applied as long as the wavelength spectrum distributions of signal and decoy states are different. The effect of this attack varies based on the discernibility of different states. However, this attack can be extended to other imperfections with distinguishable decoy states, like the polarization, temporal shape, etc. by replacing the WDM with appropriate devices.

\section{\label{numer}Numerical simulations}

\begin{figure}[tp]
  \includegraphics[width=0.43\textwidth]{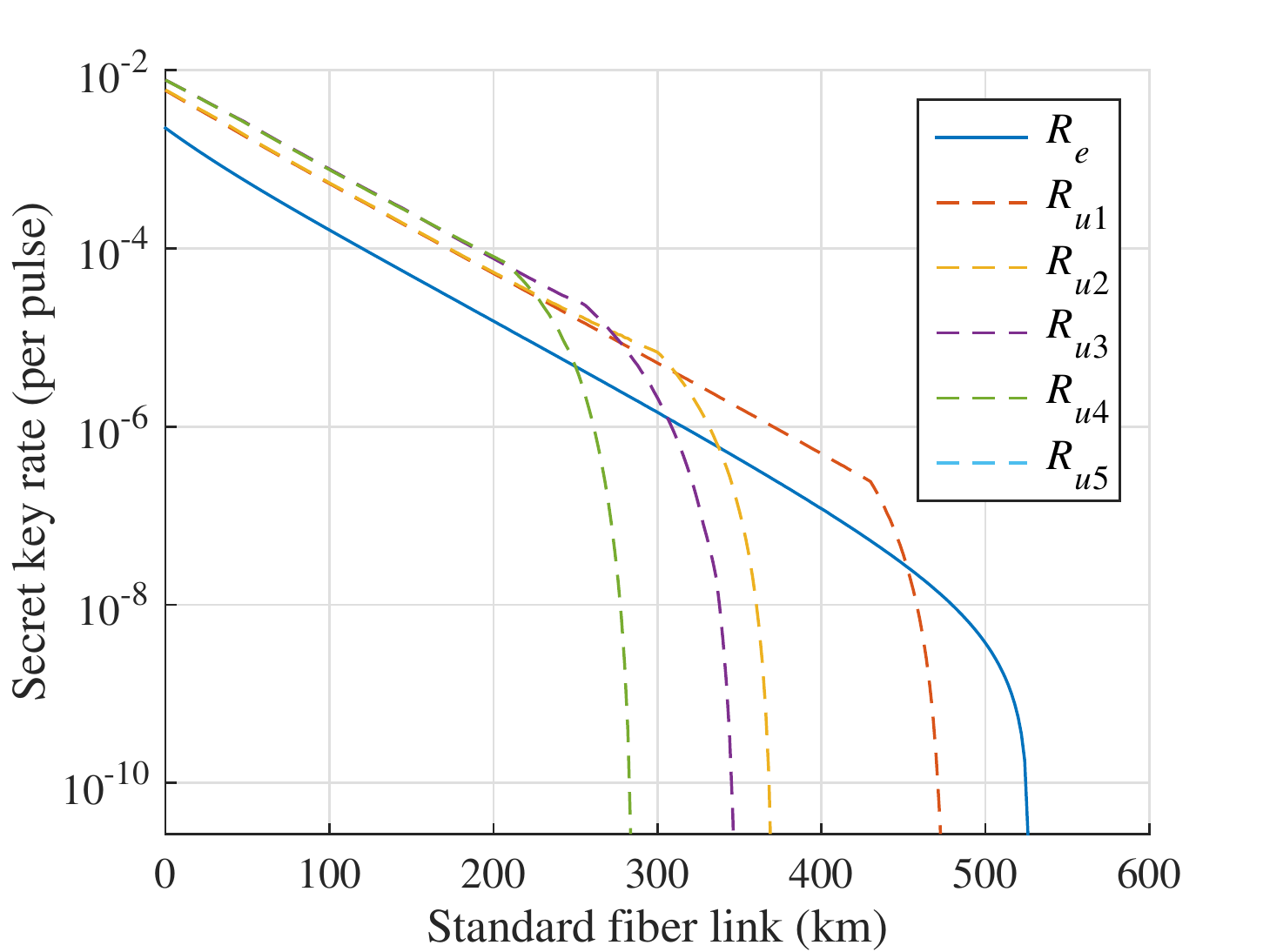}
  \caption{\label{fig4} The estimated lower bound of secret key rate $R_e$ and upper bound of real secure key rates $R_{ui}$ in logarithmic scale versus tranmission distance (between Alice and Bob) under passive frequency shift attack, where $i\in\{1,2,3,4,5\}$, with experimental parameters listed in Table \ref{tab1}, \ref{tab2} and \ref{tab3}. The solid line corresponds to the etimated key rate $R_e$ which is the same as that without attack, while the dashed lines represent the upper bound of real secure key rate $R_{ui}$. And the real secure key rate $R_{u5}$ not shown in the figure is identically 0.}
\end{figure}

We numerically simulate the behaviour of SNS protocol under passive frequency shift attack in this section. There are nine parameters obtained by statistic in practical systems before AOPP, including $n_{\alpha\beta}$, $n_{\Delta^+}^R$, $n_{\Delta^-}^L$, $n_t$ and $E_z$, where $n_t = n_{\rm sig}+n_{\rm err}$ is the length of raw keys, $\alpha \beta \in S= \{vv, va, av, vb, bv\}$. Here, $E_z=n_{\rm err}/n_t$, $n_{\rm sig}$ and $n_{\rm err}$ is the number of right and wrong raw bits, respectively. Under passive frequency shift attack, the parameters can be simulated as discussed in Appendix \ref{appnu}. The promotion to AOPP is trival as the attack is not applied.

For simulation purposes, the experimental parameters listed in Tables \ref{tab1} and \ref{tab2} are taken according to the SNS experiment \cite{chen2020RN13} with a little modification. Then we simulate the normal secret key rate without frequency shift attack, the key rate under Alice and Bob's estimation and the upper bound of secure key rate under frequency shift attack with AOPP and finite-key effects by selecting five groups of reasonable parameters about $r_{\beta|\alpha}$ listed in Tables \ref{tab3} following the principle that the difference between signal states (strong decoy states) and weak decoy states is significant while small between signal and strong decoy states.

In Fig. \ref{fig4}, the estimated key rate $R_e$ under passive frequency shift attack represented by the solid line is the same as that without attack which means this attack will not be detected by Alice and Bob with the list of experimental parameters in Table \ref{tab1}, \ref{tab2} and \ref{tab3}. In comparison, the dashed lines represent the upper bound of secure key rates $R_u$ under frequency shift attack. Denote the the upper bound of secure key rate under parameters in Group $i$ (G$i$) in Table \ref{tab3} as $R_{ui}$, where $i\in\{1,2,3,4,5\}$. Compared $R_{u1}$ with $R_{u2}$, we can find that the difference between weak decoy states and signal states (strong decoy states) affects the effects of attack significantly i.e. the communication distance is limited from 472 km to 368 km when the difference changes from 10 times to 20 times. The negative effects will be more greater when $r_{z|a}:r_{a|a}:r_{b|a}$ is constant but the absolute values increase by comparing $R_{u2}$ with $R_{u3}$. The secure distance will be limited to shorter when the difference becomes larger by contrasting $R_{u2}$ and $R_{u4}$. Besides, Alice and Bob could not distribute keys when Eve can distinguish weak decoy states correctly with parameters of Group 5. On the one hand, if Alice and Bob attach importance to these side channels and know Eve's action, this attack will limit the communication distance. Note that the distance and key rate will be more pessimistic since the key rates $R_{ui}$ is only the upper bound. Otherwise, the secret key bits will be insecure over long distance especially. For example, the key bits are all insecure when the estimated key rate is $10^{-8}$ per pulse at 479 km (an acceptable value at long distance).

In this attack, we have assumed that Eve intercepts photons in the order of $T_z$, $T_a$ and $T_b$, but we find there is no obvious difference when changing this order through numerical simulations which can be understood in this way that only the photons are secure with $ {\rm SPD1} \vee {\rm SPD2} \vee {\rm SPD3} = 0 $, i.e. only the pulses that are not detected by Eve in all internals $T_\alpha$ are secured, or there will be no difference in any order when $r_{\alpha|\beta}=0$ with $\alpha \neq \beta$.

\section{\label{diss}Discussion}

This eavesdropping attack proposed above is a passive attack harnessing the side channels and hard to be detected. To guarantee security in practical systems with side channels, the first potential way is to
improve experimental techniques or modulation methods to restrain side channels but may be unavoidable when one is closed, but another appears. The second alternative is to develop mathematical models in theory to maximize the secure key rates under attacks, like the loss-tolerant method \cite{tamaki2014RN69,pereira2019RN65,mizutani2019RN62,navarrete2020RN400,pereira2020RN255} but needs an accurate characterization of real apparatuses. It may be an ongoing search for side channels to guarantee the practical security of QKD systems.

\section{\label{conl}Conclusion}

The goal of QKD at present is to provide long-distance and high-speed key distribution, which will induce side channels inevitably. Increasing repetition rate and narrowing pulses to improve speed will make the pulses complex and distinguishable, like the frequency, polarization, temporal shape, and so on. Any small imperfections may be exploited and enhanced utilizing channel loss by Eve, especially at long distance. Therefore, it is necessary to pay more attenuation to the practical security of TF-QKD systems.

In this paper, we have investigated and tested the side channels with external modulation which is required in those TF-QKD protocols with post-phase compensation, like SNS TF-QKD \cite{wang2018RN22} and PM TF-QKD \cite{ma2018RN56}. Based on this, we propose a complete and undetected eavesdropping attack named passive frequency shift attack on SNS protocol which can be applied once there are differences between different states in frequency domain and can be extended to other imperfections with distinguishable decoy states. Normally, Alice and Bob could estimate the lower bound of secret key rate correctly no matter what Eve does. But this estimation is not accurate once Eve's operation on signal and decoy states is different which may cause insecure bits when the upper bound of secure key rate is lower than the estimated lower one. According to the numerical results, Eve can get full information about the secret key bits at long distance if Alice and Bob neglect this distinguishability. For example, the key bits are all insecure when the estimated key rate is $10^{-8}$ per pulse at 479 km under the five selected groups of parameters. As there is a variety of potentially exploitable loopholes at source, our results emphasize the practical security of the light source. It is a constant search to build hardened implementations of practical QKD systems.

\begin{acknowledgments}

This work is supported by National Key Research and Development Program of China (Grant No. 2020YFA0309702) and National Natural Science Foundation of China (Grants N0. 61605248, No. 61675235 and No. 61505261).

\end{acknowledgments}

\appendix

\section{\label{proto}SNS TF-QKD protocol}

We make a review of the SNS protocol and the key rate formula with AOPP method and finite-key effects \cite{wang2018RN22,xu2020RN23,jiang2020RN17} in this section.

(1) \textbf{Preparation and measurement.} At any time window $i$, Alice (Bob) randomly determines whether it is a signal window or a decoy window with probabilities $p_z$ and $p_x=1-p_z$. If it is a signal window, Alice (Bob) sends a phase-randomized coherent state with intensity $\mu_z$ and denotes it as 1 (0), or a vacuum state $|0\rangle$ and denotes it as 0 (1) with probabilities $p_{z1}=1-p_{z0}$ and $p_{z0}$, seperately. If it is a decoy window, Alice (Bob) sends a phase-randomized coherent state $|\sqrt{\mu_a} e^{i \theta_A} \rangle $, $|\sqrt{\mu_b} e^{i \theta_A'} \rangle $ or $|0\rangle$ ($|\sqrt{\mu_a} e^{i \theta_B} \rangle $, $|\sqrt{\mu_b} e^{i \theta_B'} \rangle $ or $|0\rangle$) with probabilities $p_a$, $p_b$ and $p_v=1-p_a-p_b$, where $\mu_a<\mu_b$. The third party, Chrelie, renamed as Eve is supposed to perform interferometic measurements on the incoming pulses and announce the results.

(2) \textbf{Different types of time windows.} Suppose Alice and Bob repeat the above process $N$ times, then they announce their signal windows and decoy windows through public channels. If both Alice and Bob determine a signal window, it is a $Z$ window. And the effective events in $Z$ windows are defined as one-detector heralded events no matter which detector clicks. Alice and Bob will get two raw $n_t$-bit strings $Z_A$ and $Z_B$ according to effective events in $Z$ windows. Note that the phase-randomized coherent state of intensity $\mu$ is equivalent to a probabilistic mixture of different photon-number states $\sum_{k=0} ^\infty \frac{e^{-\mu}\mu^k}{k!} |k\rangle \langle k|$. Therefore, we can define $Z_1$ windows as a subset of $Z$ windows when only one party determines to send and she (he) actually sends a single-photon state $|1\rangle$. The bits from effective $Z_1$ windows are regarded as untagged bits by the tagged model \cite{inamori2007RN271}. Then the intensity of pulses would be announced to each other expect the intensity in $Z$ windows. If both commit to a decoy window, it is an $X$ window. Alice and Bob also announce their phase information $\theta_A$, $\theta_B$ when they choose the same intensity $\mu_a$ in an $X$ window denoted as an $X_a$ window. And if only one detector clicks in $X_a$ windows with phases satisfying 
\begin{eqnarray}
    |\theta_A-\theta_B-\varphi_{AB}| \leq \Delta/2
    \label{eq1}
\end{eqnarray} 
or
\begin{eqnarray}
    |\theta_A-\theta_B-\pi-\varphi_{AB}| \leq \Delta/2,  
    \label{eq2}
\end{eqnarray}
it is an effective event in $X_a$ windows. All effective events in $X_a$ windows can be divided into two subsets as $C_{\Delta^+}$ and $C_{\Delta^-}$ according Eq.~\ref{eq1} and Eq.~\ref{eq2}, respectively. And the number of events in $C_{\Delta^+}$ and $C_{\Delta^-}$ can be defined as $N_{\Delta^+}$ and $N_{\Delta^-}$. Here, $\varphi_{AB}$ is set properly to obtain a satisfactory key rate which will be different over time due to phase drift and can be obtained with reference pulses. In the following, we will omit the phase drift without loss of generality and set $\varphi_{AB}=0$.

(3) \textbf{Parameter estimation.} They can estimate parameters, including the bit-flip error rate of the raw bits $E_Z$, the lower bound of untagged bits $\underline{n}_1$ (or the lower bound of the counting rate $\underline{s}_1$ equivalently) and the upper bound of the phase-flip error rate of untagged bits $\overline{e}_1^{ph}$. The bit-flip error rate $E_Z$ can be obtained by error test, $\underline{s}_1$ and $\overline{e}_1^{ph}$ can be estimated with decoy state method as follows.

Denote $\rho_v=|0\rangle \langle 0|$, $\rho_a=\sum_{k=0}^\infty e^{-\mu_a} \mu_a^k /k! |k\rangle \langle k|$ and $\rho_b=\sum_{k=0}^\infty e^{-\mu_b} \mu_b^k /k! |k\rangle \langle k|$, where $\rho_a$ and $\rho_b$ are density operators of the phase-randomized coherent states used in $X$ windows in which the phase is not announced. Let $N_{\alpha \beta}$ be the number of intsnces Alice sends state $\rho_\alpha$ and Bob sends state $\rho_\beta$ and $n_{\alpha \beta}$ be the number of corresponding one-detector heralded events, where $\alpha \beta \in S= \{vv, va, av, vb, bv\}$. Thus, the counting rate can be defined as $S_{\alpha \beta}=n_{\alpha \beta}/N_{\alpha \beta}$. And $\underline{s}_1$ can be estimated with decoy-state method as \cite{yu2013RN76,yu2019RN24}
\begin{eqnarray}
    \underline{s}_1 \geq && \frac{1}{2\mu_a \mu_b(\mu_b-\mu_a)} [\mu_b^2 e^{\mu_a} (S_{va}+S_{av}) \nonumber\\
    &&- \mu_a^2 e^{\mu_b} (S_{vb}+S_{bv}) -2(\mu_b^2-\mu_a^2) S_{vv}].
\end{eqnarray}

Denote the bit-flip errors in $C_{\Delta^+}$ ($C_{\Delta^-}$) as the effective events when the right (left) detector clicks and its total number as $n_{\Delta^+}^R$ ($n_{\Delta^-}^L$). The bit-flip error rate in $C_\Delta=C_{\Delta^+} \bigcup C_{\Delta^-}$ can be shown as
\begin{eqnarray}
    T_\Delta =\frac{n_{\Delta^+}^R + n_{\Delta^-}^L}{N_{\Delta^+} + N_{\Delta^-}}.
\end{eqnarray}
Therefore $\overline{e}_1^{ph}$ can be estimated with decoy-state method as \cite{wang2018RN22,yu2019RN24}
\begin{eqnarray}
    \overline{e}_1^{ph} \leq \frac{T_\Delta- 1/2 e^{-2\mu_a} S_{vv}}{2\mu_a e^{-2\mu_a} \underline{s}_1}.
\end{eqnarray}

(4) \textbf{Key rate formula.} With these quantities, the final key length can be expressed as \cite{tomamichel2012RN298,wang2018RN22}
\begin{eqnarray}
    R =&& 2 p_{z0}(1-p_{z0}) \mu_z e^{-\mu_z} \underline{s}_1 [1-H(\overline{e}_1^{ph})] \nonumber\\
    &&- n_t f H(E_Z)/N.
\end{eqnarray}
where $N_f$ is the number of final bits, $H(x)=-x {\rm log}_2 x-(1-x) {\rm log}_2 (1-x)$ is the binary entropy function, and $f$ is the error correction efficiency factor. 

(5) \textbf{AOPP method.} AOPP method \cite{xu2020RN23,jiang2020RN17} is a pre-error correction process on raw strings $Z_A$ and $Z_B$ proposed to improve the direct tranmission key rate. In AOPP method, Bob randomly select two unequal bits as pairs and will obtain $n_p={\rm min}(n_{t0},n_{t1})$ pairs, where $n_{t0}$ ($n_{t1}$) is the number of bits 0 (1) in raw string $Z_B$. There will be only two types of pairs can be survived when Alice make exactly the same or opposite decision as Bob for two bits, and denote the number as $n_{vd}$ or $n_{cc}$, respectively. Therefore, the bit error after AOPP is shown as
\begin{eqnarray}
    E_Z'=\frac{n_{vd}}{n_{cc}+n_{vd}}.
\end{eqnarray}
The lower bound of the number of untagged bits is
\begin{eqnarray}
    \underline{n}_1' = n_p \frac{\underline{n}_1^0}{n_{t0}} \frac{\underline{n}_1^1}{n_{t1}},
\end{eqnarray}
where $\underline{n}_1^0$ and $\underline{n}_1^1$ is the lower bound of untagged bits when they make the opposite decision and obtain bits 0 and 1, correspondingly. And the phase-flip error rate changed into $ {\overline{e}'}_1^{ph} =2\overline{e}_1^{ph}(1-\overline{e}_1^{ph}) $. Besides, finite-key effects should be considered in practical systems using Chernoff bound \cite{chernoff1952RN144,curty2014RN100} and the parameters can be estimated as $n_1' = \varphi^L (\underline{n}_1') $ and $e_1'^{ph} = \varphi^U ( \underline{n}_1' \overline{e}_1'^{ph} ) / \underline{n}_1'$. Finally, the improved key length can be shown as \cite{jiang2020RN17,xu2020RN23,tomamichel2012RN298}
\begin{eqnarray}
    N_f'=&& n_1'[1- H({e'}_1^{ph})]-n_t' f H(E_Z')-{\rm log}_2 \frac{2}{\varepsilon_{cor}} \nonumber\\
    &&-2{\rm log}_2 \frac{1}{\sqrt{2} \varepsilon_{PA} \hat\varepsilon}.
\end{eqnarray}

\section{\label{detail}Details of the upper bound of secure key rate}

To obtain the upper bound of secure key rate, we should consider the ideal situation, i.e. the upper bound of the number of secure single photons and the lower bound of phase-flip error rate without finite-key effects. 

Ideally, the single photons which could be used to distill secure key bits (i.e. the upper bound of secure untagged bits) after AOPP can be shown as
\begin{eqnarray}
  n_{1,{\rm sec}} = n_p \frac{n_{1s}^0}{n_{t0}} \frac{n_{1s}^1}{n_{t1}},
\end{eqnarray}
where
\begin{eqnarray}
  n_{1s}^0 = n_{1s}^1 = N p_z^2 p_{z0} (1-p_{z0}) p_{zk} u_z \eta_{zk} e^{-u_z \eta_{zk}}.
\end{eqnarray}
In the rest, we will analyze this inherent phase-flip error rate from the perspective of virtual protocol. In the virtual protocol \cite{wang2018RN22}, Alice and Bob will prepare an extended state
\begin{eqnarray}
  | \varPsi \rangle = \frac{1}{\sqrt{2}} ( e^{i\delta_B} |01\rangle \otimes |01\rangle + e^{i\delta_A} |10\rangle \otimes |10\rangle ),
\end{eqnarray}
with restriction of Eq. \eqref{eq1} or \eqref{eq2} which is equivalent to the state $ [ |01\rangle \langle 01| \otimes |01\rangle \langle 01| + |10\rangle \langle 10| \otimes |10\rangle \langle 10| ] / 2 $ when Alice and Bob measure ancillary photons in photon-number basis in advance. Those states left of $\otimes$ are real states which will be sent to Charlie, while the right are local ancillary states with bit value encoded. Local state $|0\rangle$ corresponds to a bit 0 (1), and state $|1\rangle$ corresponds to a bit 1 (0)  for Alice (Bob). In order to obtain lower bound of phase-flip error rate, consider the ideal situation in which the phase shift can be compensated perfectly which is equivalent to no phase shift. After interference and detection, the local states change into
\begin{eqnarray}
  \rho_l = \frac{1}{2} [ |\varphi_1\rangle \langle \varphi_1 | +  |\varphi_2\rangle \langle \varphi_2 |],
\end{eqnarray}
where
\begin{eqnarray}
    |\varphi_1\rangle &&= [ |01\rangle + e^{i\delta} |10\rangle ]/ \sqrt{2},\nonumber\\
    |\varphi_2\rangle &&= [ |01\rangle - e^{i\delta} |10\rangle ]/ \sqrt{2}, 
\end{eqnarray}
with $\delta = \delta_A - \delta_B$. When the local ancillary states measured virtually with basis $|\Phi^0 \rangle=[ |01\rangle + |10\rangle ]/ \sqrt{2}$ and $|\Phi^1 \rangle=[ |01\rangle - |10\rangle ]/ \sqrt{2}$, the phase-flip error rate before AOPP can be shown as
\begin{eqnarray}
    e_{\rm in}' = \frac{e_{\rm in}^0+e_{\rm in}^1}{2},
\end{eqnarray}
where $e_{\rm in}^0$ and $e_{\rm in}^1$ are phase-flip error rate when $\delta_A$ and $\delta_B$ satisfy Eq. \eqref{eq1} and \eqref{eq2}, respectively,
\begin{eqnarray}
    e_{\rm in}^0 &&= {\rm Tr} [ |\Phi^1 \rangle \langle \Phi^1 | \rho_l ],\nonumber\\
    e_{\rm in}^1 &&= {\rm Tr} [ |\Phi^0 \rangle \langle \Phi^0 | \rho_l ].
\end{eqnarray}
On average, the inherent phase-flip error will be $\overline{e}_{\rm in}'= \int_{-\pi/M}^{\pi/M} e_{\rm in}' dM\delta/2\pi $ and is approximately equal to 0.0032 when $M=16$. Thus, we can obtain the inherent phase-flip error rate with AOPP method as $e_{\rm in} = 2\overline{e}_{\rm in}'(1-\overline{e}_{\rm in}')$.

\section{\label{appnu}Details of numerical simulations}

Under passive frequency shift attack, the parameters obtained by statistic can be shown as follows
\begin{eqnarray}
    n_{\rm sig} =&& 4Np_z^2 p_{z0}p_{z1} \sum_{\gamma \in M} p_{z\gamma} \bigl[\overline{p}_d e^{-\frac{\mu_{z\gamma}}{2}} - \overline{p}_d^2 e^{-\mu_{z\gamma}}\bigr],\\
    n_{\rm err} =&& 2Np_z^2 \Big[p_{z1}^2  \sum_{\gamma_1, \gamma_2 \in M} p_{z\gamma_1} p_{z\gamma_2} \bigl[ -\overline{p}_d^2 e^{-(\mu_{z\gamma_1}+\mu_{z\gamma_2})} \nonumber\\
    && +\overline{p}_d  e^{ -\frac{\mu_{z\gamma_1} + \mu_{z\gamma_2}} {2} } I_0 ( \sqrt{\mu_{z\gamma_1} \mu_{z\gamma_2}}) \bigr] + p_{z0}^2 p_d \overline{p}_d \Bigr], \\
    n_{\alpha v} =&& 2N_{\alpha v} \sum_{\gamma \in M} p_{\alpha \gamma} [\overline{p}_d e^{-\mu_{\alpha\gamma}/2} - \overline{p}_d^2 e^{-\mu_{\alpha\gamma}}],\\
    n_{v \beta} =&& 2N_{v \beta} \sum_{\gamma \in M} p_{\beta \gamma} [\overline{p}_d e^{-\mu_{\beta \gamma}/2} - \overline{p}_d^2 e^{-\mu_{\beta \gamma}}],\\
      n_{vv} =&& 2N_{vv}p_d \overline{p}_d,
\end{eqnarray}
where $p_d$ is the dark count rate and $\overline{p}_d = 1-p_d$, and $\mu_{\beta \gamma}=\mu_\beta \eta_{\beta\gamma}$. Here, $\beta \in \{a,b,z\}$ and $\gamma \in M$. Note that the intensities of state $|e^{i\theta_A} \sqrt{\mu_a \eta_{a \gamma_1}} \rangle$ and $|e^{i\theta_B} \sqrt{\mu_a \eta_{a \gamma_2}} \rangle$ from Alice and Bob in $X_a$ windows may be different after passive frequency shift attack where $\gamma_1$, $\gamma_2 \in M$, but this does not mean it could not cause right detection. After interference, the intensity of left and right detectors will be
\begin{eqnarray}
      \mu_l (\gamma_1,\gamma_2) =&& \frac{1}{2} \Bigl[ \mu_{a\gamma_1}+ \mu_{a\gamma_2} +2\sqrt{\mu_{a\gamma_1} \mu_{a\gamma_2}} {\rm cos}\delta \Bigr], \\
      \mu_r (\gamma_1,\gamma_2) =&& \frac{1}{2} \Bigl[ \mu_{a\gamma_1}+ \mu_{a\gamma_2} -2\sqrt{ \mu_{a\gamma_1} \mu_{a\gamma_2}} {\rm cos}\delta \Bigr],
\end{eqnarray}
where $\delta=\theta_B-\theta_A$. We can see that the difference of two output intensities does not determined by the distinction of two input intensities but phase difference. Then the number of error events in $C_{\Delta^\pm}$ can be shown as
\begin{eqnarray}
      n_{\Delta^+}^R &&=  N_{\Delta^+} \sum_{\gamma_1,\gamma_2 \in W} p_{a\gamma_1} p_{a\gamma_2} \Big[ - \overline{p}_d^2 e^{-\mu_{a\gamma_1}-\mu_{a\gamma_2}} \nonumber\\
      && + \overline{p}_d \int_{-\Delta/2}^{\Delta/2} e_d e^{-\mu_r(\gamma_1,\gamma_2)} + \overline{e}_d e^{-\mu_l(\gamma_1,\gamma_2)} d\frac{\delta}{\Delta}  \Bigr],
\end{eqnarray}
where $e_d$ is the misalignment-error probability and $\overline{e}_d=1-e_d$. Similarly, we can obtain $n_{\Delta^-}^L$.

\end{document}